\documentstyle[multicol,epsf,aps]{revtex}

\begin{document}
\draft
\title{Chiral-glass transition and replica symmetry breaking
  of a three-dimensional Heisenberg
  spin glass}
\author{K. Hukushima}
\address{ISSP, Univ. of Tokyo, Roppongi, Minato-ku, Tokyo 106-8666,
  Japan}
\author{H. Kawamura}
\address{Faculty of Engineering and Design, Kyoto Institute of 
Technology,
  Sakyo-ku, Kyoto 606-8585,  Japan}
\date{\today}
\maketitle
\begin{abstract}
Extensive equilibrium Monte Carlo simulations are performed for a 
three-dimensional Heisenberg spin glass with
the nearest-neighbor Gaussian coupling to investigate its
spin-glass and chiral-glass orderings. The occurrence of a
finite-temperature chiral-glass transition 
without the conventional spin-glass order is established.
Critical exponents characterizing the transition are
different from those of the standard Ising spin glass.
The calculated overlap distribution suggests the appearance of
a peculiar type of replica-symmetry breaking
in the chiral-glass ordered state.
\end{abstract}
\begin{multicols}{2}
\narrowtext

While experiments have provided convincing
evidence that spin-glass (SG) magnets exhibit 
an equilibrium  phase transition 
at a finite temperature,
the true nature of the experimentally observed SG
transition and that of the low-temperature SG phase still remain
open problems\cite{Review}. 
A simple Ising model has widely been used as a ``realistic'' SG
model in the studies, {\it e.g.\/}, of
the critical
properties of the SG transition, or of the 
issue of whether the SG state exhibits a spontaneous
replica-symmetry breaking (RSB).
One should bear in mind, however, that 
the magnetic interactions in  many SG
materials are nearly isotropic, being 
well described by an isotropic
Heisenberg model, in the sense that the magnetic anisotropy is
considerably weaker than the exchange interaction.  
In apparent contrast to experiments,
numerical simulations have indicated
that the standard spin-glass order 
occurs only at zero temperature in the three-dimensional (3D)
Heisenberg SG\cite{OYS,Kawamura92,Kawamura95,Kawamura98,Matsubara91}. 
Although the magnetic anisotropy inherent  
to real materials
is often invoked to explain this apparent discrepancy with 
experiments,
it still remains puzzling that  no 
detectable sign of
Heisenberg-to-Ising crossover has been observed in experiments
which is usually expected to occur if the observed
finite-temperature transition 
is caused by the weak magnetic anisotropy\cite{Review,OYS}. 

In order to solve this apparent puzzle, a  chirality mechanism of
experimentally observed spin-glass transitions
was proposed by one of the authors\cite{Kawamura92}. This scenario
is based on the assumption 
that an isotropic 3D Heisenberg SG
exhibits a finite-temperature 
{\it chiral-glass\/} transition without the conventional spin-glass 
order, in which only spin-reflection 
symmetry is broken with
preserving spin-rotation symmetry. Chirality is an Ising-like
multispin variable representing the
sense or the handedness of the noncollinear spin structures induced
by spin frustration. In this scenario, all
essential features of many of real SG transitions and 
SG ordered states
should be determined by the properties of
the chiral-glass 
transition and of the chiral-glass state
{\it in the fully isotropic system\/}, while the  role of the
magnetic anisotropy  is secondary which
``mixes'' the spin and the chirality and
``reveals'' the chiral-glass transition as an anomaly
in experimentally accessible quantities.

Numerical studies on the 3d XY spin glasses 
have given strong support to the occurrence of a
finite-temperature chiral-glass  transition \cite{KT91,XYCG,Maucourt}. 
In the Heisenberg case, 
while previous numerical works 
agreed in that
the standard SG order occurred only at $T=0$,
the question whether there really
occurs a finite-temperature chiral-glass
order has remained inconclusive\cite{Kawamura92,Kawamura95}.
Very recently, an off-equilibrium 
Monte Carlo simulation by one of the authors
has given  evidence for the occurrence of a 
finite-temperature chiral-glass order in the 3D Heisenberg SG
\cite{Kawamura98}. 
However, the full critical properties 
of the chiral-glass transition
as well as the properties of the chiral-glass ordered state itself,
particularly the question of the possible RSB,  still
remains largely unclear.
In this Letter, we perform extensive {\it equilibrium\/}
Monte Carlo simulations
of a 3D Heisenberg SG in order to determine the 
detailed static and dynamic 
critical properties 
and to clarify the nature of the 
chiral-glass ordered state.

Our model is the classical Heisenberg
model on 
a simple cubic lattice with $N=L^3$ spins
defined by the Hamiltonian,
\begin{equation}
  \label{model}
  {\cal H}  = -\sum_{\langle ij\rangle}J_{ij}{\bf S}_i\cdot {\bf S}_j,
\end{equation}
where ${\bf S}_i$ 
=$(S_i^x,S_i^y,S_i^z)$ 
is a three-component 
unit vector, and 
the sum runs over all nearest-neighbor pairs. 
The interactions $J_{ij}$ are random Gaussian variables 
with zero mean and variance $J$.  
The local chirality at the $i$-th site and in the $\mu $-th 
direction, 
$\chi _{i\mu }$, is defined for 
{\it three\/} neighboring spins 
by the scalar\cite{OYS,Kawamura95},
\begin{equation}
  \label{localc}
  \chi_{i \mu}={\bf S}_{i+\hat{e}_\mu}\cdot({\bf S}_i\times
{\bf S}_{i-\hat{e}_\mu}),
\end{equation}
where $\hat {\bf e}_\mu \ (\mu =x,y,z)$ 
denotes a unit lattice vector along the
$\mu\/$-axis.

Monte Carlo simulation is performed based on the exchange MC 
method, sometimes called ``parallel tempering'',
which turns out to be an efficient method for thermalizing systems
exhibiting slow dynamics\cite{HN}. 
By making use of this method, 
we have succeeded in equilibrating
the system down to the temperature considerably  lower than those
attained in the previous simulations\cite{OYS,Kawamura95}.
We run two independent sequences of  systems 
(replica 1 and 2) in parallel, and
compute an overlap between the chiral variables in the two
replicas,
\begin{equation}
  q_{\chi} = \frac{1}{3N}\sum_{i\mu}\chi_{i\mu}^{(1)}\chi_{i\mu}^{(2)}.
\end{equation}
In terms of this chiral overlap $q_{\chi} $, 
the Binder ratio of the chirality
is calculated by
\begin{equation}
  \label{Binder}
  g_{\rm CG}(L)=\frac{1}{2}\left(3-\frac{[\langle q_\chi^4\rangle]}
    {[\langle  q_\chi^2\rangle]^2} \right), 
\end{equation}
where $\langle\cdots\rangle$ represents the thermal average 
and $[\cdots ]$ 
represents the average over bond disorder. 
For the  Heisenberg spin,
one can introduce an appropriate Binder ratio in terms of 
a tensor overlap $q_{\mu \nu}\ (\mu ,\nu =x,y,z)$
which has $3^2=9$ independent components\cite{Kawamura95}, 
\begin{equation}
q_{\mu\nu}=\frac{1}{N}\sum_i S_{i\mu}^{(1)}S_{i\nu}^{(2)}  \ \ \ 
(\mu,\nu=x,y,z),
\end{equation}
via the relation,
\begin{equation}
  \label{BinderSG}
  g_{\rm  SG}(L)=
\frac{1}{2}\left(11-9\frac{\sum_{\mu,\nu,\delta,\rho}[\langle
    q_{\mu\nu}^2q_{\delta\rho}^2\rangle]}{(\sum_{\mu,\nu} [\langle
    q^2_{\mu\nu}\rangle])^2} \right).
\end{equation}
The lattice sizes studied are
$L=6,8,10,12$ and $16$ with periodic boundary conditions.
Equilibration is checked by monitoring the 
stability of the results
against at least three-times longer runs for a subset of samples.
Sample average is taken over 1500 ($L=6$), 1200 ($L=8$), 640 ($L=10$),
296 ($L=12$) and 136 ($L=16$) independent bond realizations.
Note that in the exchange MC simulations the data at different
 temperatures are correlated.  
Error bars are estimated from statistical fluctuations over the bond
 realizations.

The size and temperature dependence 
of the Binder ratios of the spin and of the chirality,
$g_{{\rm SG}}$ and $g_{{\rm CG}}$, are shown in
Fig.~1(a) and (b), respectively. As can be seen from Fig.~1(a),
$g_{{\rm SG}}$ constantly decreases with increasing $L$ 
at all temperatures studied,
suggesting that 
the conventional spin-glass order occurs only at
zero temperature, consistent with the previous results
\cite{OYS,Kawamura92,Kawamura95,Kawamura98,Matsubara91}.
Fig.~1(a) reveals that $g_{{\rm SG}}$ 
for larger lattices ($L=10,12,16$) exhibits an anomalous bending 
around $T/J\simeq 0.15$, suggesting a change in the ordering
behavior  in this temperature range. 
As can be seen from Fig.~1(b), 
the curves  of $g_{{\rm CG}}$ for different $L$ 
tend to merge for larger $L$
in the temperature range where the curves of $g_{{\rm SG}}$ exhibit
an anomalous bending.
Furthermore, on increasing $L$,
the merging points gradually move toward higher temperatures, 
suggesting that
the chiral-glass transition indeed occurs at a
finite temperature.
Since $g_{{\rm CG}}$ for different $L$ do not
cross here at $g_{\rm CG}>0$, however, 
it is not necessarily easy to unambiguously locate the chiral-glass
transition point, 
or even to completely rule out the possibility of only
a zero-temperature transition with rapidly
growing ({\it e.g.\/}, exponentially growing) 
correlation length. Meanwhile, the calculated
$g_{{\rm CG}}$ shows a negative dip whose depth gradually increases
with increasing $L$, whereas its position gradually
shifts toward lower temperature. 
Note that, 
if the systems would not exhibit any finite-temperature transition,
$g_{{\rm CG}}(T; L\rightarrow \infty )$
should be equal to zero at any $T>0$ and to unity at $T=0$
(for the 
nondegenerate ground state as expected for the present Gaussian
coupling).  Since $g_{{\rm CG}}(T; L\rightarrow \infty )$ is
a single-valued function of $T$,
the existence of a  
negative dip growing with $L$ 
hardly reconciles with the
absence of a finite-temperature transition 
so long as this tendency persists, and suggests
the chiral-glass transition occurring at
$T=T_{{\rm CG}}>0$ at which $g_{{\rm CG}}$ takes a
{\it negative\/} value unlike the standard
cases of the 3D Edwards-Anderson
(EA) Ising model or the infinite-range Sherrington-Kirkpatrick
(SK) model.

\begin{figure}[h]
\label{fig:Binder}
\epsfxsize=\columnwidth\epsfbox{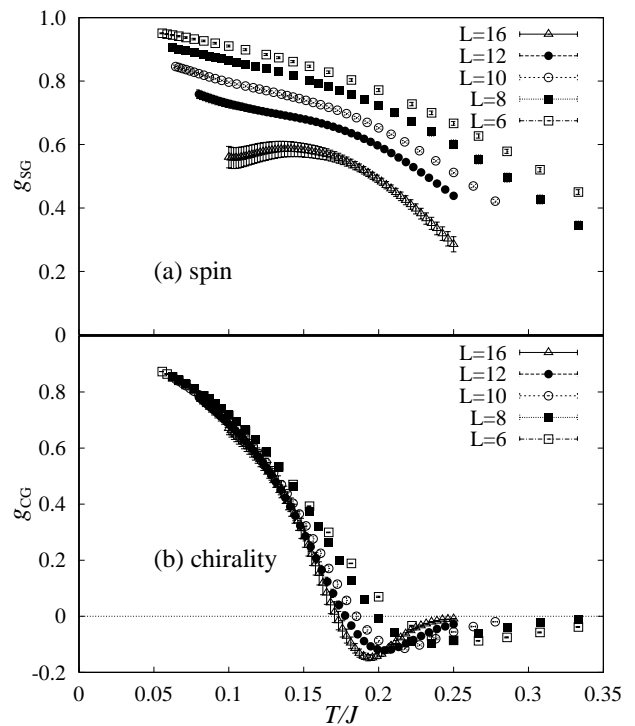}
\caption{
Temperature and size dependence of the Binder ratios
of the spin (a) and  of the chirality (b). 
}
\end{figure}

More unambiguous estimate of $T_{{\rm CG}}$ can be obtained from
the equilibrium dynamics of the model. 
Thus, we calculate both the spin and chirality autocorrelation 
functions defined by
\begin{eqnarray}
C_s(t) & = & {1\over N}
\sum _i[\langle{\bf S}_i(t_0)\cdot {\bf S}_i(t+t_0)\rangle], \\
C_\chi (t) & = & {1\over 3N}\sum _{i,\mu }[\langle\chi _{i\mu }(t_0) 
\chi _{i\mu }(t+t_0)\rangle], 
\end{eqnarray}
where MC simulation  is performed 
according to the standard heat-bath updating here.
The starting spin configuration at $t=t_0$ is taken from 
the equilibrium spin configuration
generated in our exchange MC runs.

Monte Carlo time dependence of the
calculated $C_s(t)$ and $C_\chi (t)$ 
are shown in Fig.~2 on log-log plots
at  several temperatures for $L=16$.
We found no significant difference in the data of L=12 and 16,
    and the finite-size effect is negligible in our time window.
As can be seen from Fig.~2(a), $C_s(t)$ 
shows a downward curvature at all
temperature studied, suggesting an exponential-like decay
characteristic of the disordered phase, consistently with the
absence of the standard spin-glass order. In sharp contrast to this,
$C_\chi (t)$ shows either a downward curvature 
characteristic of the disordered phase, or an upward curvature
characteristic of the long-range ordered phase, 
depending on whether the temperature is higher or lower than
$T/J\simeq 0.16$, 
while just at $T/J\simeq 0.16$ the linear behavior corresponding to
the power-law decay 
is observed: See Fig.~2(b).
Hence, our dynamical data indicates that the chiral-glass
order without the standard spin-glass order 
takes place at $T_{{\rm CG}}/J=0.160\pm 0.005$,  
below which a finite chiral
EA order parameter $q_{{\rm CG}}^{{\rm EA}}>0$ develops.
From the slope of the data at $T=T_{{\rm CG}}$,
the exponent $\lambda $ characterizing the power-law decay of 
$C_\chi (t)\approx t^{-\lambda }$ is estimated to be
$\lambda =0.193\pm 0.005$. The estimated $T_{{\rm CG}}$ is in good
agreement with the previous estimate of Ref.~\cite{Kawamura98}, 
$T_{{\rm CG}}/J=0.157\pm 0.01$.

\begin{figure}[h]
\label{fig:eq-dynamics}
\epsfxsize=0.98\columnwidth\epsfbox{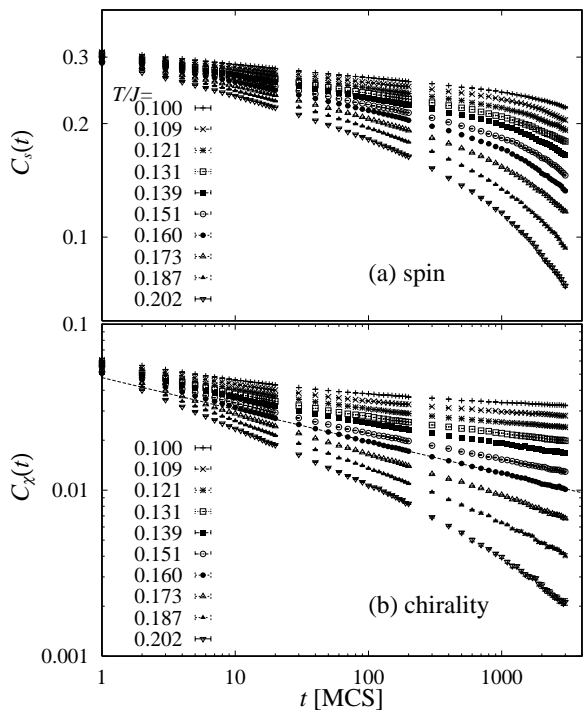}
\caption{
Log-log plots of the 
time dependence of the equilibrium
spin (a) and chirality (b) autocorrelation 
functions 
at several temperatures.
The lattice size
is $L=16$ averaged over 64 samples. In (b), the best straight-line 
fit is obtained at $T/J=0.16$,  represented by the
broken line. 
}
\end{figure}

The behavior of the chiral-glass order parameter, or the
associated chiral-glass susceptibility $\chi _{{\rm CG}}
=3N[\langle q_\chi ^2\rangle]$, 
turns out to be consistent with this. 
In the inset of Fig.~3, 
we show the reduced chiral-glass susceptibility 
$\tilde \chi _{{\rm CG}}\equiv \chi _{{\rm CG}}/\bar \chi^4$,
normalized by the amplitude of the local chirality $\bar\chi ^2 
\equiv (1/3N)\sum _{i, \mu}[\langle\chi _{i\mu}^2\rangle]$, 
versus the reduced temperature 
$t\equiv\mid(T-T_{{\rm CG}})/T_{{\rm CG}}\mid$ on a log-log plot. 
From the asymptotic slope of the 
data,
the susceptibility exponent is estimated to be 
$\gamma _{{\rm CG}}=1.5\pm ^{0.3}_{0.1}$. 
Note that the estimated susceptibility
exponent is significantly smaller than that of the standard
3D Ising EA model, $\gamma \simeq  4$ [1d].

If we combine the present estimate of $\gamma _{{\rm CG}}$ with
the estimate of $\beta _{{\rm CG}}$ from the off-equilibrium 
simulation of Ref.~\cite{Kawamura98} and
use the scaling relations, 
various chiral-glass exponents can be estimated to be
$\alpha \simeq -1.7$, $\beta_{{\rm CG}}\simeq 1.1$,
$\gamma_{{\rm CG}}\simeq 1.5$,  $\nu_{{\rm CG}}\simeq 1.2$ and
$\eta _{{\rm CG}}\simeq 0.8$. The dynamical exponent 
is estimated to be $z_{{\rm CG}}\simeq 4.7$ 
by using the estimated value of $\lambda $
and the scaling relation $\lambda =\beta _{{\rm CG}}/z_{{\rm CG}}\nu 
_{{\rm CG}}$.
While the dynamical exponent $z_{{\rm CG}}$ comes rather 
close to the $z$ of the 3D EA model,
the obtained static exponents differ significantly
from those of the 3D Ising EA model
$\beta\simeq 0.55$,
$\gamma\simeq 4.0$,  $\nu\simeq 1.7$ and $\eta \simeq -0.35$ [1d],
suggesting that the universality class of the chiral-glass
transition of the 3D Heisenberg spin glass 
differs from that of the standard 3D Ising spin glass. 
Possible long-range and/or many-body nature of the chirality-chirality
interaction might be the cause of this deviation.

\begin{figure}[h]
\label{fig:G-para}
\epsfxsize=\columnwidth\epsfbox{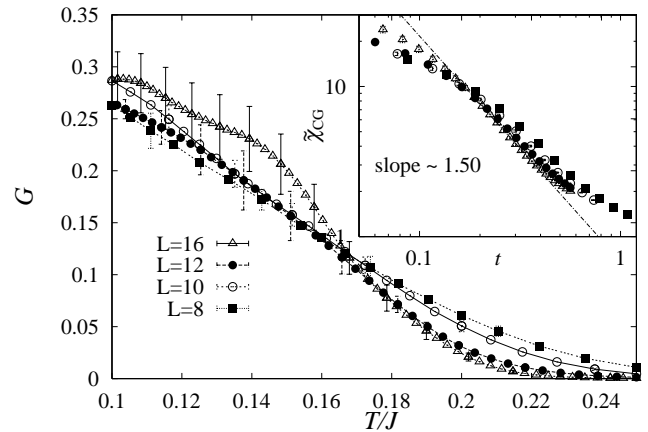}
\caption{Temperature and size dependence of the $G$ parameter of
 the chirality
 defined by
 Eq. (9). The inset 
 represents a log-log 
 plot of the reduced
 chiral-glass susceptibility versus the reduced temperature
 $t\equiv \mid (T-T_{{\rm CG}})/T_{{\rm CG}}\mid $.
}
\end{figure}

Further evidence of a phase transition is obtained from the
behavior of the $G$ parameter of the chirality defined by
$$ G(L) = \frac{[\langle q_\chi^2\rangle^2]-[\langle q_\chi^2\rangle]^2}
  {[\langle q_\chi^4\rangle]-[\langle q_\chi^2\rangle]^2}.\eqno(9)$$
While this quantity was originally introduced to represent the
non-self-averaging character of the system\cite{lackSA}, Bokil {\it et al\/}
argued that it was not necessarily so\cite{comment}. Still, a crossing of 
$G(L)$ for different $L$, if it occurs, can be used to identify the  
transition\cite{comment}. As shown in Fig.~3, for $T>T_{{\rm CG}}$
$G(L)$ decreases with increasing $L$ tending to zero, 
while for $T<T_{{\rm CG}}$ it tends to increase with $L$, thus
lending further support to the existence of a phase transition
at $T=T_{{\rm CG}}$.

\begin{figure}[h]
\label{fig:pofq}
\epsfxsize=\columnwidth\epsfbox{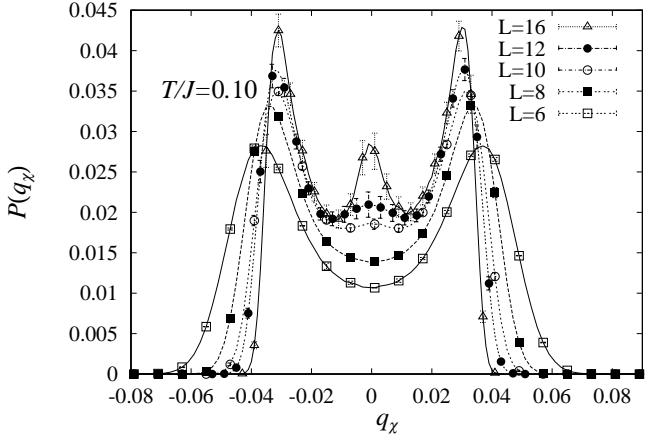}
\caption{
Chiral-overlap distribution function 
below $T_{{\rm CG}}$. The temperature is $T/J=0.1$.
}
\end{figure}
In Fig.~4, we display the distribution function of the chiral-overlap 
defined by 
$P(q'_\chi )=[\langle\delta (q_\chi -q'_\chi )\rangle]$
%
calculated at a temperature $T/J=0.1$, well below the 
chiral-glass transition temperature. 
The shape of the calculated $P(q_\chi )$ is
somewhat different from the  one 
observed in the standard Ising-like models such as the
3D EA model or the mean-field
SK model. As usual, 
$P(q_\chi )$ has standard ``side-peaks''
corresponding to the Edwards-Anderson order parameter 
$\pm q_{{\rm CG}}^{{\rm EA}}$, which 
grow and sharpen with increasing $L$. The extracted value of
$\pm q_{{\rm CG}}^{{\rm EA}}$ coincides with that evaluated from
the relaxation of $C_\chi (t)$.
In addition to the side peaks, 
an unexpected  
``central peak'' at $q_\chi =0$ shows up for larger $L$, 
which also grows and sharpens with increasing $L$. This latter
aspect, {\it i.e.\/}, the existence of a central peak, 
is a peculiar feature of
the chiral-glass ordered state never
observed in 
the EA or SK models. Since we 
do not find any sign of a first-order transition such as a
discontinuity in the energy, the specific heat nor the order parameter
$q_{{\rm CG}}^{{\rm EA}}$, 
this feature is likely to be related to a 
nontrivial structure in the phase space associated with the
chirality. 
We note that this peculiar feature
is  reminiscent of the behavior characteristic of  some mean-field
models showing the so-called {\it 
one-step\/} RSB[1b].
Indeed, the existence of a negative
dip in the Binder ratio $g_{{\rm CG}}$  and the absence of a crossing of  
$g_{{\rm CG}}$ at $g_{\rm CG}>0$ are    
consistent with the occurrence of such one-step-like RSB\cite{Potts}.
Our data of $P(q_\chi )$ are 
also not incompatible with
the existence of a continuous plateau between 
[$-q_{{\rm CG}}^{{\rm EA}},q_{{\rm CG}}^{{\rm EA}}$]
in addition to
the delta-function peaks.
According to the chirality mechanism, such novel one-step-like RSB should
be realized in the {\it spin} ordering of real Heisenberg-like spin glasses.
We note that, if $P(q_\chi)$ has a nontrivial structure as suggested 
from Fig.~4, the denominator of Eq.~(9) should  
remain nonzero at $0<T<T_{\rm CG}$. 
Then, our data of $G(L)$ in Fig.~3 
indicates  that the chiral-glass state is non-self-averaging.

In summary, spin-glass and chiral-glass orderings of the 
3D Heisenberg
SG are studied by Monte Carlo simulations. Our
observation both on statics and dynamics  strongly suggests the
existence of a stable chiral-glass phase at finite temperatures 
without the
conventional spin-glass order.
This  fact strengthens the plausibility of the
chirality mechanism for experimentally
observed spin-glass transitions. 
The nature of the chiral-glass
ordered state as well as of the critical phenomena 
are different
from those of the 3D Ising SG, with strong similarities to the 
system showing the
one-step-like RSB, while its exact nature and 
physical origin have remained to be understood.

The numerical calculation was performed on the Fujitsu VPP500 at the
supercomputer center, ISSP, University of Tokyo, on the HITACHI
SR-2201 at the supercomputer center, University of Tokyo, and on
the CP-PACS computer at the Center for Computational Physics, 
University  of Tsukuba.

\end{multicols}
\end{document}